India's Progress in Space Exploration and International Legal Challenges in Meeting Goals within International Space Boundaries: A Review

Jayanthi Vajiram [1], Utkarsh Maurya [2] Negha Senthil [3]

[1]Vellore Institute of Technology, Chennai, jayanthi.2020@vitstudent.ac.in(Teaching Research Associate)
[2,3]Vellore Institute of Technology, utkarsh.maurya2019@vitstudent.ac.in, negha.2019@vitstudent.ac.in.

**Abstract:**

India's journey towards space has been nothing short of remarkable. It began with the establishment of the Indian Space Research Organization (ISRO) in 1969, with the aim of harnessing space technology for national development. Since then, India has made significant strides in space exploration, satellite technology, and space research. One of the major milestones in India's space journey was the launch of the first Indian satellite, Aryabhata, on April 19, 1975. This marked India's entry into the exclusive club of space-faring nations. The success of Aryabhata paved the way for further advancements in space technology and inspired a generation of scientists and engineers. Another significant achievement came in 1984 when India launched its first indigenous satellite, INSAT-1A. This satellite played a crucial role in revolutionizing communication and broadcasting in the country. It provided services like television, telecommunication, meteorology, and disaster warning systems. India's most notable achievement in space exploration came in 2008 with the successful launch of the Chandrayaan-1 mission. This mission aimed to explore the moon and gather valuable data about its surface and composition. Chandrayaan-1 not only confirmed the presence of water molecules on the lunar surface but also discovered evidence of widespread volcanic activity in the moon's past. In 2013, India achieved another significant milestone with the launch of the Mars Orbiter Mission (MOM), also known as Mangalyaan. It was India's first interplanetary mission and made India the first country to successfully enter Mars orbit on its first attempt. The MOM mission not only made India the fourth country to reach Mars but also achieved this feat at a remarkably low cost, showcasing India's capabilities in space technology frugality. India's space program has also been focused on satellite technology and its applications for national development. The Indian National Satellite (INSAT) system has played a crucial role in providing various essential services to the country. These include communication, broadcasting, weather forecasting, remote sensing, and resource mapping. The INSAT system has significantly contributed to the socio-economic development of India, bridging the digital divide and enabling connectivity in remote areas. Furthermore, India's space program has also been dedicated to scientific research and exploration. The Indian Space Observatory ASTROSAT, launched in 2015, is the country's first multi-wavelength space telescope. It has been instrumental in studying celestial objects, understanding black holes, and exploring the mysteries of the universe. Collaboration has also been an integral part of India's space journey. ISRO has collaborated with several international space agencies, including NASA, ESA, and Ros cosmos, on various space missions and research projects. These collaborations have facilitated the exchange of knowledge, resources, and expertise, leading to significant advancements in space science and technology. In recent years, India has set ambitious goals for its space program. The Chandrayaan-2 mission, launched in 2019, aimed to land a rover on the moon's surface and further explore its uncharted regions. Though the lander did not achieve a soft landing, the mission still provided valuable data and insights for future lunar exploration endeavors. India's space program is also gearing up for its first manned mission, Gaganyaan, scheduled for 2022. This mission will make India the fourth country to send humans to space. It is a testament to India's advancements in space technology and its commitment to pushing boundaries and exploring the unknown.

**Introduction:**

The growth of India's space program has not been limited to scientific and technological achievements. It has also had significant socio-economic impacts. The space industry has created numerous job opportunities for scientists, engineers, technicians, and support staff. It has also spurred the growth of related industries, such as manufacturing, telecommunications, and aerospace engineering and has contributed to the overall economic growth of the country. India's journey towards space has not been without challenges. The nature of space exploration and research involves inherent risks and complexities. However, India has consistently shown resilience, determination, and a spirit of innovation in overcoming these challenges. Looking ahead, India's space program is poised for further growth and achievements. With plans for more ambitious missions, increased focus on research and development, and collaborations with international partners, India aims to establish itself as a prominent player in the global space community. In conclusion, India's journey towards space has been a testament to its scientific prowess, technical capabilities, and unwavering commitment to national development. From early milestones to groundbreaking missions, India's space program continues to advance the frontiers of science, technology, and exploration. It has not only brought immense pride to the nation but has also laid the foundation for a brighter and more technologically advanced future. India's foray into space exploration has gained momentum in recent years, with notable successes such as the Mars Orbiter Mission (Mangalyaan) and the Chandrayaan missions. These achievements have showcased India's technological capabilities and ambitions in the realm of space exploration. However, as India pursues its space objectives, it confronts legal complexities and challenges that arise from international space boundaries and the need to comply with existing laws and agreements. Indian scientists have made significant contributions to the field of space exploration. The Indian Space Research Organization (ISRO) has achieved numerous milestones under the leadership of prominent scientists. Their efforts have resulted in successful missions such as the Mars Orbiter Mission (Mangalyaan), which made India the first Asian country to reach Mars. Another noteworthy achievement is the Chandrayaan-2 mission, which included a lunar orbiter, lander, and rover. These missions demonstrate the dedication and expertise of Indian scientists in advancing space research and exploration.

**Literature review**

The several key areas of concern and legal challenges that India encounters in meeting its targets within international space boundaries. These include issues related to national sovereignty in outer space, space debris management, satellite launches and licensing, intellectual property rights, liability for damage caused by space activities, and compliance with international treaties and agreements.

National Sovereignty in Outer Space: India, like other nations, asserts its sovereignty and jurisdiction over objects launched into space. However, questions arise regarding the extent to which a nation's laws and regulations apply in outer space and how they intersect with international legal principles. Space Debris Management: With an increasing number of satellites and space missions, the issue of space debris becomes critical. India faces the challenge of managing space debris generated by its own activities and complying with international guidelines for debris mitigation. Satellite Launches and Licensing and regulating satellite launches and space activities requires compliance with national laws and international obligations, including the registration of space objects and adhering to frequency coordination regulations. Intellectual Property Rights and the protection of technological innovations and advancements. Balancing national interests, commercial interests, and international obligations becomes crucial in this context. Liability for damage caused by space objects requires careful consideration, particularly in cases of accidents, collisions, or satellite malfunctions. Compliance with International Treaties and agreements governing space activities, such as the Outer Space Treaty, Liability Convention, and Registration Convention. Ensuring compliance with these obligations while pursuing national space objectives presents legal challenges that need to be addressed. This paper adopts a literature review approach to gather and analyze existing research papers, articles, and relevant literature related to India's space journey and the legal complications it faces within international space boundaries. The research sources include scholarly papers, that provide insights into India's space program and the international legal frameworks governing space activities. Figure 1 explains the frame work of this paper.

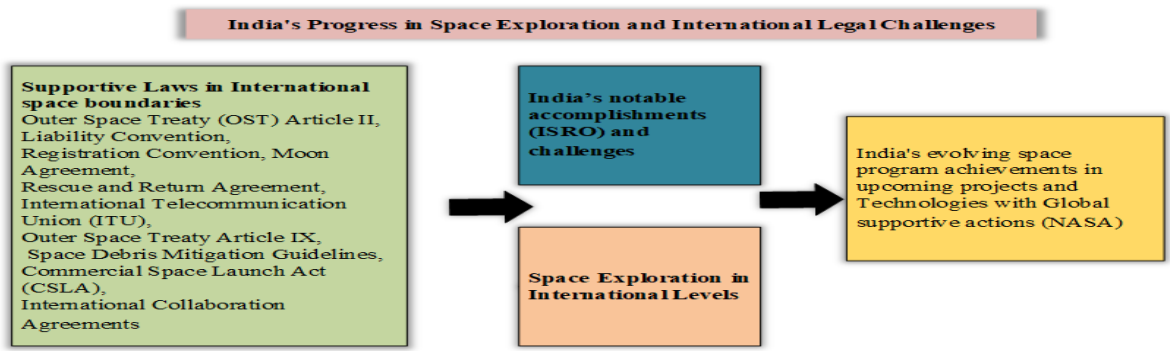

Figure 1: India's progress in Space Exploration and International Legal Challenges

**India's notable accomplishments (ISRO) and challenges:**

1. Mars Orbiter Mission (Mangalyaan): Launched in November 2013, Mangalyaan is India's first interplanetary mission, and it made India the first Asian country to reach Mars orbit and the fourth space agency globally to do so. The mission was a great success, demonstrating India's capabilities in deep space exploration.

2. Chandranan Missions: ISRO has launched two lunar missions - Chandrayaan-1 and Chandrayaan-2. Chandrayaan-1, launched in 2008, was India's first lunar mission and made significant discoveries, including the detection of water molecules on the Moon's surface. Chandrayaan-2, launched in 2019, aimed to explore the Moon's south pole region and its findings further enriched our understanding of the lunar surface.

3. Navigation with Indian Constellation (NavIC): NavIC is India's regional satellite navigation system, designed to provide accurate position information to users in India and the surrounding region. It is an indigenous counterpart to the GPS system of the United States and became operational in 2018.

4. Satellite Launch Capability: ISRO has developed various launch vehicles, including the Polar Satellite Launch Vehicle (PSLV) and the Geosynchronous Satellite Launch Vehicle (GSLV), which have been instrumental in launching numerous satellites for various purposes, such as communication, remote sensing, weather monitoring, and scientific research.

5. Reusable Launch Vehicle (RLV) Technology: ISRO has been working on developing reusable launch vehicle technology, aiming to reduce the cost of access to space by reusing certain components of the launch vehicle.

6. Gaganyaan Mission: ISRO is working on its first human spaceflight program called Gaganyaan. The mission aims to send Indian astronauts (Gagannauts) into space onboard the spacecraft and bring them back safely. The first crewed mission is planned for the future.

7. Interplanetary Exploration: Besides Mars, ISRO has ambitious plans for exploring other planets and celestial bodies in the solar system. It has proposed missions to Venus (Shukrayaan) and also plans for a mission to study the Sun (Aditya-L1).

8. International Collaborations: ISRO actively engages in international collaborations and partnerships, both in space research and commercial satellite launches.

9. GSLV Mk III (LVM3): The Geosynchronous Satellite Launch Vehicle Mark III is India's heaviest and most powerful launch vehicle. It has the capacity to carry heavier payloads to higher orbits and is crucial for future manned space missions.

10. Navigation with Indian Constellation (NavIC): NavIC is India's regional satellite navigation system, similar to GPS. It provides accurate position information service to users in India and the surrounding region.

11. International Collaborations: ISRO has collaborated with various international space agencies and organizations for joint missions, satellite launches, and technology development.

12. AstroSat: Launched in 2015, AstroSat is India's first dedicated multi-wavelength space observatory, enabling observations in ultraviolet, visible, and X-ray wavelengths.

13. Launch Services for Commercial Payloads: ISRO has established a reputation as a reliable and cost-effective launch service provider, attracting commercial customers from around the world.

ISRO has earned recognition and respect globally for its cost-effectiveness, technical capabilities, and numerous successful space missions These achievements showcase India's commitment to space exploration, technology development, and scientific research. ISRO's cost-effective approach, innovative technologies, and ambitious goals have earned it global recognition in the space community.

**Space Exploration in International Levels**

H.R.4752, known as "The Space Development and Settlement Act," was introduced in the 114th Congress in 2016. The bill aims to promote the development of space resources and the establishment of settlements beyond Earth. It emphasizes the utilization of space resources for economic, scientific, and technological purposes. The bill encourages the federal government to facilitate commercial space activities, provide incentives for private investment, and create a regulatory framework for space settlement. It envisions fostering partnerships between government and private sectors to enable sustainable space exploration and colonization. The bill's objective is to establish a legal foundation that supports the expansion of human presence in space and the exploration of outer space resources [1]. The paper by Stephen Gorove in 1969, discusses the interpretation of Article II of the Outer Space Treaty, the paper focuses on legal implications related to the Outer Space Treaty's prohibition of national appropriation of celestial bodies. The author explores the treaty's language, historical context, and intent to decipher its impact on sovereignty claims in outer space. Gorove examines key arguments surrounding property rights and resource utilization, emphasizing the need for international cooperation and legal clarity to avoid conflicts. The paper addressing the challenges of ensuring peaceful and equitable use of outer space resources [2]. Antonella Forganni's 2017 paper, "The potential of space tourism for space popularization: An opportunity for the E.U. Space Policy?" examines the role of space tourism in popularizing space exploration. Published in Space Policy, Volume 41, the paper discusses how space tourism could contribute to the European Union's space policy. It explores the potential benefits of space tourism for public engagement and awareness of space activities. The author considers economic, regulatory, and technological aspects of integrating space tourism into the EU's space strategy. The paper emphasizes the significance of public interest, private sector involvement, and international collaboration in realizing the potential of space tourism. It underscores the importance of policy considerations to harness the benefits of space tourism for both popularization and space industry growth [3]. In 2012, James E. Dunstan was set to deliver remarks at the Competitive Enterprise Institute's (CEI) event titled "Property Rights in Space." The event aimed to discuss property rights issues in outer space exploration. The event was hosted by Tech Freedom, and Dunstan's remarks were expected to provide insights into the legal and policy aspects of property rights in space activities. The event's focus was likely on the evolving legal framework surrounding space resource utilization and property ownership in space. CEI's commitment to exploring property rights in the context of space exploration reflects the ongoing discussions and debates about the legal and ethical considerations of space activities and resource utilization [4]. The "Declaration of Legal Principles Governing the Activities of States in the Exploration and Use of Outer Space," issued in 1962, outlines fundamental principles for states engaging in space exploration. Published by the United Nations Office for Outer Space Affairs (UNOOSA), the document sets guidelines for international cooperation, peaceful use of outer space, and equitable sharing of benefits. It emphasizes that outer space is open to exploration by all countries for peaceful purposes and prohibits placing nuclear weapons or other weapons of mass destruction in orbit. The declaration underscores the need to avoid harmful contamination of celestial bodies and promotes the use of space for the common interest of all humanity. It serves as a foundational document in shaping international space law and cooperation efforts [5]. In Bryon C. Brittingham's 2010 paper titled "Does the world need new space law" published in the Oregon Review of International Law, the author raises the question of whether new legal frameworks are required for governing activities in outer space, the paper examines the evolving challenges posed by advancements in space technology and exploration. Brittingham delves into the adequacy of existing space law in addressing issues such as commercial space activities, resource utilization, and space debris. The author likely analyzes whether international law adequately addresses these concerns or if new regulations are necessary. The paper contributes to the ongoing discourse about the necessity of updating space law to accommodate the changing landscape of space activities [6]. Erazem Bohinc's 2013 paper "International Space Law: legal aspects of exploiting outer space" explores legal dimensions related to utilizing outer space. Published in Nova Gorica, the paper delves into international space law, discussing the legal framework for activities beyond Earth. The author likely examines topics such as space exploration, resource exploitation, satellite operations, and potential conflicts. The paper might address the rights and responsibilities of nations, private entities, and international cooperation in space endeavors. Bohinc's work contributes to understanding the legal challenges and opportunities in the field of space exploration and exploitation. The paper's focus on the legal aspects reflects the importance of robust legal frameworks for fostering peaceful and cooperative activities in space [7].

**International Legal Challenges and supportive laws within International Space Boundaries:**

Navigating international legal challenges within the context of international space boundaries presents a complex endeavor. As nations strive to achieve their space-related objectives, several legal considerations come into play.

Resource Utilization of space resources, such as minerals or energy sources, raises questions about property rights and equitable sharing among nations. International agreements must be developed to ensure fair access and prevent conflicts over resource exploitation. Sovereignty and Non-Appropriation enshrined in the Outer Space Treaty prevents nations from claiming celestial bodies as their own. Balancing this with sovereignty rights becomes crucial, especially as space activities become more ambitious and diverse. Environmental concerns in Space activities can generate debris and contamination that affect the space environment. Developing legal frameworks for responsible and sustainable practices becomes essential to mitigate the impact on space ecosystems. Peaceful use and security concerns while upholding the peaceful use of outer space requires international cooperation and robust legal mechanisms. Liability and Accountability crafts the legal frameworks to address liability issues in a rapidly evolving space environment is essential for resolving disputes. Cybersecurity and data Sharing rely heavily on communication and data exchange, safeguarding against cyber threats and establishing protocols for secure data sharing are critical legal aspects.

The growing involvement of the private sector in space activities necessitates legal clarity regarding ownership, liability, and regulations to ensure responsible and commercially viable practices. International Collaborative projects, like the International Space Station, require intricate legal agreements to govern contributions, responsibilities, and intellectual property rights among participating nations. Developing legal frameworks to protect intellectual property rights generated through space exploration, technology development, and research is essential to encourage innovation. As nations pursue ambitious space goals, the need for effective global governance mechanisms becomes more evident. International cooperation, agreements, and forums are essential for harmonizing legal norms and ensuring a peaceful and collaborative space environment.

Outer Space Treaty (OST) Article II, Liability Convention, Registration Convention, Moon Agreement, Rescue and Return Agreement, International Telecommunication Union (ITU), Outer Space Treaty Article IX, Space Debris Mitigation Guidelines, Commercial Space Launch Act (CSLA), International Collaboration Agreements. These supportive laws and sections contribute to the development of a cohesive legal framework that facilitates international cooperation, responsible exploration, and equitable utilization of outer space resources while addressing the challenges

**Recent technologies used in space:**

a. CubeSats and Small Satellites: Miniaturized satellites, such as CubeSats, have gained popularity for their cost-effectiveness and versatility. They are used for various purposes, including Earth observation, communication, and scientific research.

b. Reusable Rockets: Technologies like SpaceX's Falcon 9 have introduced reusability to space launches, significantly reducing launch costs and making space access more economical.

c. Electric Propulsion: Electric propulsion systems, such as ion and Hall-effect thrusters, provide efficient and prolonged thrust for space vehicles, enabling deep space exploration and station-keeping. Advancements in solar panel technology enhances power generation in space, supporting longer missions and energy-intensive instruments.

e. Advanced Imaging Instruments: High-resolution cameras and imaging spectrometers enable detailed observation of celestial bodies, contributing to planetary exploration and astronomical research.

f. Autonomous Navigation and AI: Autonomous navigation systems and artificial intelligence enhance spacecraft's ability to navigate, perform complex tasks, and adapt to changing conditions in space.

g. 3D Printing in Space: Additive manufacturing allows astronauts to produce tools, spare parts, and even structures in space, reducing the need for resupply from Earth.

h. Interplanetary Communication: Deep space communication networks like NASA's Deep Space Network and the European Space Agency's (ESA) Estrack provide continuous communication with spacecraft across vast distances.

i. In-Situ Resource Utilization (ISRU): Technologies for extracting and utilizing resources from celestial bodies, such as water ice on the Moon, could support future long-duration missions and colonization.

j. Next-Generation Telescopes: Telescopes like the James Webb Space Telescope (JWST) offer improved resolution and observation capabilities, allowing us to explore distant galaxies and unravel cosmic mysteries.

k. Space-Based Internet: Initiatives like SpaceX's Starlink aim to create global internet coverage using constellations of low Earth orbit satellites.

l. Lunar and Martian Landers: Advancements in landing technology are enabling precise and safe landings on the Moon and Mars for exploration and resource assessment.

m. Quantum Communication: Quantum communication experiments in space offer the potential for ultra-secure data transmission between Earth and satellites.

n. Life Support Systems: Developments in life support technologies are crucial for sustaining astronauts on long-duration missions, including Mars expeditions.

o. These recent technologies demonstrate the rapid progress and innovation in space exploration, pushing the boundaries of our understanding and opening new frontiers for scientific discovery and human presence beyond Earth.

**India's evolving space program achievements in upcoming projects and Global supportive actions (NASA) Upcoming Projects:**

Gaganyaan: India's manned space mission aims to send astronauts into space by 2022. It demonstrates India's capability in human spaceflight. Chandrayaan-3, a dedicated lander mission to explore the Moon's south pole, following the partial success of Chandrayaan-2. Aditya-L1 mission aims to study the Sun and its outermost layer, the corona, contributing to solar physics research.

**Global Supportive Actions (NASA Collaboration):**

Deep Space Network (DSN) communication capabilities have been enhanced through collaboration with NASA's DSN, benefiting missions like Mangalyaan. NASA's LRO (Lunar Reconnaissance Orbiter) provided critical data for India's Chandrayaan-1 mission, enhancing lunar exploration. NASA contributed with its Lunar Crater Observation and Sensing Satellite (LCROSS) data to support Chandrayaan-2's objectives. India Joint Exploration with US Working Group promotes collaboration on future lunar and Mars exploration missions. Global Space Situational Awareness Network (GSSAP), led by the U.S., for sharing data on space object tracking and situational awareness. These achievements, upcoming projects, and collaborative efforts with NASA reflect India's commitment to space exploration, technology development, and international cooperation in advancing our understanding of space and its potential benefits for humanity.

The Indian space program, managed by the Indian Space Research Organization (ISRO), has been known for its cost-effectiveness. The funding for India's space activities primarily comes to ISRO through Government Budget Allocation, Project-Specific Funding, Commercial Activities, International Collaboration, Cost-Effective Approach, Private Sector Participation.

**Conclusion:**

India's evolving space program reflects three key priorities. Firstly, as the program advances in complexity, it embraces space exploration as a natural progression. Second, while some missions lack immediate developmental gains, they enhance the country's visibility and attractiveness for international partnerships. Third, these missions have spurred technological innovations, with advancements in deep space communication attributed to collaborations like those with NASA. Notably, India's initial Moon mission and the Mangalyaan deep space communication endeavor received assistance from NASA. Overall, these missions and technological strides bolster India's influence in shaping global outer space governance. India's progress in space exploration is commendable, but it is not without legal challenges. This literature review provides insights into these challenges and highlights the need for India to navigate international legal frameworks and align its space activities with global standards. By acknowledging and addressing these legal complexities, India can continue its space journey while maintaining international cooperation and compliance. . Effective international cooperation and the development of comprehensive legal frameworks are crucial to ensuring the responsible and equitable exploration and utilization of outer space.